\documentclass[conference]{IEEEtran}
\IEEEoverridecommandlockouts

\usepackage{cite}
\usepackage{amsmath,amssymb,amsfonts}
\usepackage{algorithmic}
\usepackage{graphicx}
\usepackage{textcomp}
\usepackage{xcolor}

\usepackage{siunitx}
\usepackage{hyperref}
\usepackage{tikz}

\usepackage{xcolor}

\usepackage{lipsum}
\usepackage[pscoord]{eso-pic}
\newcommand{\placetextbox}[3]{
  \setbox0=\hbox{#3}
  \AddToShipoutPictureFG*{
    \put(\LenToUnit{#1\paperwidth},\LenToUnit{#2\paperheight}){\vtop{{\null}\makebox[0pt][c]{#3}}}%
  }%
}%

\def\BibTeX{{\rm B\kern-.05em{\sc i\kern-.025em b}\kern-.08em
    T\kern-.1667em\lower.7ex\hbox{E}\kern-.125emX}}
\begin{document}
\placetextbox{0.5}{1}{This is the author's version of an article that has been published.}
\placetextbox{0.5}{0.985}{Changes were made to this version by the publisher prior to publication.}
\placetextbox{0.5}{0.97}{The final version of record is available at \href{https://doi.org/10.1109/ETFA65518.2025.11205662}{https://doi.org/10.1109/ETFA65518.2025.11205662}}%
\placetextbox{0.5}{0.05}{Copyright (c) 2025 IEEE. Personal use is permitted.}
\placetextbox{0.5}{0.035}{For any other purposes, permission must be obtained from the IEEE by emailing pubs-permissions@ieee.org.}%

\title{Wi-Fi Rate Adaptation for Moving Equipment in Industrial Environments\\

\thanks{This work was partially supported by the European Union under the Italian National Recovery and Resilience Plan (NRRP) of NextGenerationEU, partnership on ``Telecommunications of the Future'' (PE00000001 - program ``RESTART'').}
}

\author{\IEEEauthorblockN{Pietro Chiavassa}
\IEEEauthorblockA{\textit{CNR–IEIIT} \\
\textit{National Research Council of Italy}\\
Turin, Italy \\
pietrochiavassa@cnr.it}
\and
\IEEEauthorblockN{Stefano Scanzio}
\IEEEauthorblockA{\textit{CNR–IEIIT} \\
\textit{National Research Council of Italy}\\
Milan, Italy \\
stefano.scanzio@cnr.it}
\and
\IEEEauthorblockN{Gianluca Cena}
\IEEEauthorblockA{\textit{CNR–IEIIT} \\
\textit{National Research Council of Italy}\\
Turin, Italy \\
gianluca.cena@cnr.it}
}

\maketitle

\begin{abstract}
Wi-Fi is currently considered one of the most promising solutions for interconnecting mobile equipment (e.g., autonomous mobile robots and active exoskeletons) in industrial environments.
However, relability requirements imposed by the industrial context, such as ensuring bounded transmission latency, are a major challenge for over-the-air communication.
One of the aspects of Wi-Fi technology that greatly affects the probability of a packet reaching its destination is the selection of the appropriate transmission rate.
Rate adaptation algorithms are in charge of this operation, but their design and implementation are not regulated by the IEEE 802.11 standard.
One of the most popular solutions, available as open source, is Minstrel, which is the default choice for the Linux Kernel.

In this paper, Minstrel performance is evaluated for both static and mobility scenarios.
Our analysis focuses on metrics of interest for industrial contexts, i.e., latency and packet loss ratio, and serves as a preliminary evaluation for the future development of enhanced rate adaptation algorithms based on centralized digital twins.
\end{abstract}


\section{Introduction}

Wireless communication technologies are increasingly used in industrial environments to support mobility of devices, including complex machinery with moving parts.
In particular, recent versions of IEEE 802.11 \cite{IEEE802-11-2024} (Wi-Fi 4 to 7) offer high throughput at a limited cost by operating in unlicensed frequency bands with commercial equipment.
The most important requirements when dealing with control applications are dependability and responsiveness \cite{2023-TNSM-TSN}.
The former is typically characterized by the packet loss ratio (PLR), i.e., the fraction of generated packets that, in spite of retransmissions carried out by the MAC layer, never reach destination.
The second can be instead described by means of statistics on transmission latency, e.g., mean value and high percentiles.

Both these metrics are related to the probability $\epsilon$ that a frame transmission attempt may fail.
The higher $\epsilon$ (which typically varies over time and space), the higher the latency (because of retries) and the worse the PLR (because of the retry limit).
In turn, the failure probability is affected by a number of aspects like the path loss (which increases with the distance between communication endpoints, typically a wireless STA and the AP it is associated to) and the level of disturbance in the area where devices are deployed (with a clear distinction between visible and hidden interferers).

Probability $\epsilon$ also depends on the modulation and coding scheme (MCS) selected for frame transmission (which determines the data rate).
Importantly, the MCS is not necessarily the same for all packets sent by a given STA and for all the attempts performed for the same packet.
Generally speaking, the lower the data rate, the lower $\epsilon$, but specific phenomena like multipath fading may lead to unexpected behaviors.

For this reason, specific rate adaptation (RA) algorithms~\cite{Chaudhary2023} are customarily adopted, which dynamically select the optimal MCS at runtime.
A popular example is Minstrel \cite{minstrel}, which is included in the Linux Kernel.
In particular, they try to determine which MCS is expected to achieve the best performance,
using approaches that resemble reinforcement learning.
This MCS, which in theory should offer the optimal behavior in the near future, is used for \textit{exploitation}.
Contextually, \textit{exploration} is also carried out, which permits to determine if environmental conditions have changed, hence making the algorithm adaptive.
This has a cost, since some attempts are purposely made with non-optimal MCSs to probe the surrounding spectrum.
Several works in the literature show that RA is quite effective in improving  Wi-Fi communication quality~\cite{chiavassa2025suitabilitywifiinterconnectingmoving}.

A very important aspect when RA algorithms are used in contexts characterized by device mobility is that they typically rely on the statistics evaluated on transmission attempts performed in the recent past, e.g., by means of moving averages.
Their effect resembles a low-pass filter, which introduces some delay and worsens RA effectiveness in scenarios characterized by fast dynamics.
For instance, if a device is moving at high speed, the ability of Minstrel to determine the optimal MCS is noticeably impaired.

In this paper, we identified \textit{latency} as the key performance indicator for communication in industrial environments.
Then, we evaluated some relevant statistics about latency on a Wi-Fi link (mean value and percentiles) versus the distance between communication end-points, when one of these end-points is either static or moved back and forth on a linear trajectory.
Movements on this trajectory were identically repeated a large number of times, in order to collect reliable statistics.

In the next section, the methodology we used for evaluation is described, 
while in Section~\ref{sec:Results} results are presented and discussed.
Some conclusions are finally drawn in Section~\ref{sec:Conclusions}.

\section{Methodology}

The experiments we carried out are aimed at analyzing the communication quality between a Wi-Fi station (STA), referred to as the STA Under Test (SUT), and the access point (AP) to which it is associated, as a function of the distance between the two devices.
Differently from previous work~\cite{chiavassa2025suitabilitywifiinterconnectingmoving}, the objective of this study is to evaluate the effect of STA mobility, and in particular the SUT moving speed (APs are typically deployed in fixed positions),
on Minstrel performance.
Considered \textit{metrics} are the percentage of dropped packets, together with the mean and 99th percentile of latency.
All experiments are conducted via simulation using the ns-3 library.
SUT trajectory is constrained to a straight line, allowing us to simplify the analysis by reducing it to a one-dimensional problem.
However, this is not a limiting assumption, since the effects of communication quality are mainly related to the distance between the SUT and the AP, as well as between them and possible interferers.

\subsection{Network configurations}
\begin{figure}[b]
    \centerline{\includegraphics[width=0.95\columnwidth]{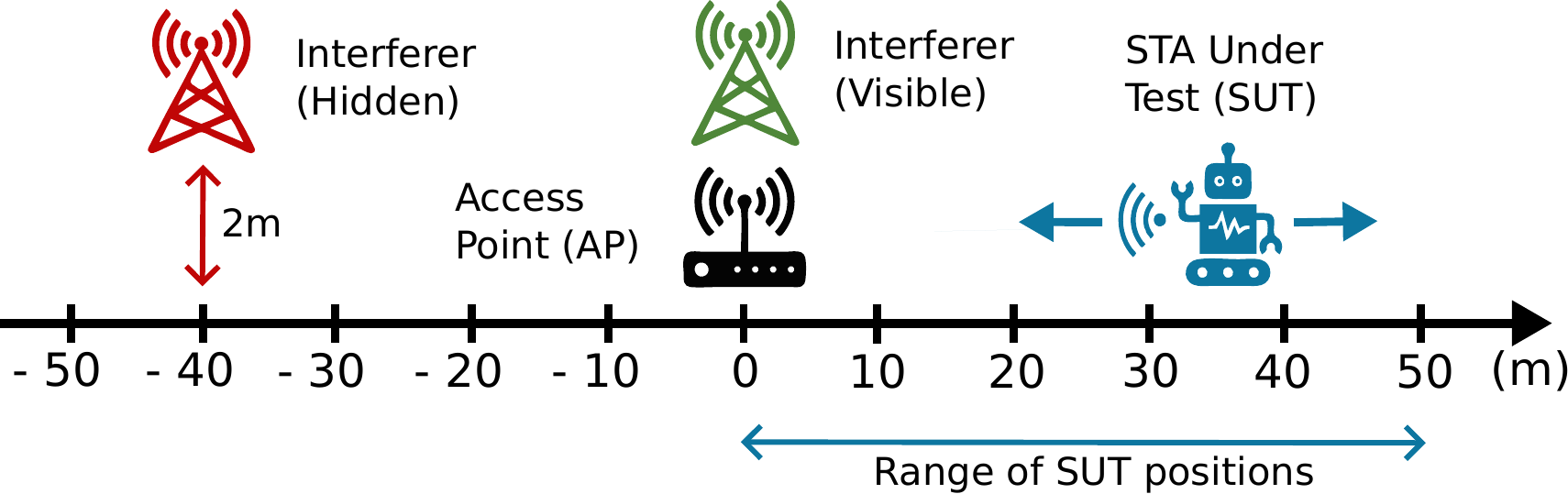}}
    \caption{Physical placement of devices (one-dimensional map).}
    \label{fig:Map}
\end{figure}

Three different \textit{network configurations} are analyzed, which are summarized in Fig.~\ref{fig:Map}.
In the simplest configuration (NO\_INT), only the AP and the SUT are present.
The AP is positioned at the origin of the Cartesian coordinate system, while the SUT can change its position $D_\mathrm{S}$ (expressed as the distance from the AP) from $D_\mathrm{S}=\SI{1}{m}$ to $D_\mathrm{S}=\SI{50}{m}$.

In the second network configuration (VISIBLE), an additional client STA, acting as an interferer and referred to as INT, is positioned at a fixed distance $D_\mathrm{I}=\SI{0}{m}$ from the AP, to which it is also associated.
To prevent the antennas of INT and the AP from occupying exactly the same position, INT was displaced by $\SI{2}{m}$ in the orthogonal direction.

In the last configuration (HIDDEN) the interfering STA is placed at a fixed distance $D_\mathrm{I}=\SI{-40}{m}$ from the AP, while maintaining its orthogonal displacement.
Unlike the VISIBLE configuration, when $D_\mathrm{S}>\SI{11}{m}$, i.e., INT and the SUT are more than $\SI{51.45}{m}$ apart, they are not able to see each other's traffic, leading to the hidden node effect.
This results in collisions between the packets they send to the AP, because both stations always sense the channel to be free.

\subsection{Simulation setup}
Both the SUT and the interfering STA run applications that generate UDP packets directed to a server installed in the AP.
The SUT traffic pattern is characterized by periodic messages, transmitted every $\SI{0.5}{s}$, to model a typical industrial device exchanging control messages.
The size of the message payload (22 Bytes) is selected to fit inside a single IEEE 802.11 PSDU.
Latency is evaluated as the time interval between the generation of the packet at the application layer and the reception of its acknowledgment.

The interfering STA, instead, generates bursty traffic, mimicking general-purpose applications that execute bandwidth-intensive operations (file transfer, multimedia streaming, etc.).
Both the spacing between bursts and their packet number are randomly selected according to exponential distributions, with an average of $\SI{250}{\mu s}$ and 100 packets, and bounded to a maximum of $\SI{10}{s}$ and 500 packets, respectively.
Packet payload is 1472 B, and packet spacing inside each burst is set to $\SI{500}{\mu s}$.

Simulated devices were configured to operate on channel $44$ in the $\SI{5}{GHz}$ frequency band 
with $\SI{20}{MHz}$ channel width.
The Wi-Fi standard selected for all simulations is IEEE 802.11a.
This is done to reduce complexity and to disable all features targeted at improving throughput, such as multiple streams, channel bonding, and frame aggregation, which are not beneficial for the SUT given the type of traffic it generates (short periodic messages).
The analysis of mechanisms introduced in Wi-Fi 6 and 7, such as trigger frames and multi-link operation (MLO), which are currently not widely supported, is left as future work.
Finally, RTS/CTS is not enabled, since the duration of the related control frames is comparable to the data frames generated by the SUT (hence, no benefits would be provided by their adoption).

For what concerns the PHY layer, it is simulated via the ns-3 \textit{SpectrumWifiPhy} class.
Channel conditions are described via the \textit{SpectrumChannel} class, using the \textit{ConstantSpeedPropagationDelayModel} and \textit{LogDistancePropagationLossModel} classes to characterize the signal propagation speed and path loss, respectively.
By using default parameters for all instantiated objects, the maximum communication range between two Wi-Fi devices is  $D_\mathrm{max}=\SI{51.45}{m}$.

\subsection{Mobility}
For each network configuration, different \textit{mobility scenarios} were analyzed.
In the simplest case (\textit{static}), independent simulations are run by positioning the SUT at fixed locations $D_\mathrm{S} \in [\SI{1}{m}, \SI{50}{m}]$, considering $\SI{1}{m}$ steps.
The duration of each simulation is set to $\SI{30000}{s}$ to obtain detailed statistics of the Minstrel behavior in static conditions.

Three mobility scenarios were considered where the SUT moves at constant speed from $D_\mathrm{S}=\SI{0}{m}$
to $D_\mathrm{S}=\SI{51}{m}$ and then comes back to $D_\mathrm{S}=\SI{0}{m}$.
In each simulation, depending on speed, this path is repeated enough times so that the SUT remains in each $\SI{1}{m}$ interval for $\SI{30000}{s}$, as for the static scenario.
Three different speeds were considered for the SUT: $\SI{1}{cm/s}$ (\textit{mobility\_slow}),  $\SI{10}{cm/s}$ (\textit{mobility\_medium}), and $\SI{1}{m/s}$ (\textit{mobility\_fast}).
This could model a heavy AGV that moves along paths in an indoor factory environment, where people are also present.
Packet transmissions on which to compute statistics are aggregated depending on the position of the SUT at the time of their initiation, rounding the value to the nearest integer (meters).
In this way, each discretized location $D_\mathrm{S} \in [\SI{1}{m}, \SI{50}{m}]$ is characterized, on average, by the same number of packet transmissions.

\section{Results}
\label{sec:Results}
\begin{figure}[t]
    \centerline{\includegraphics[width=0.9\columnwidth]{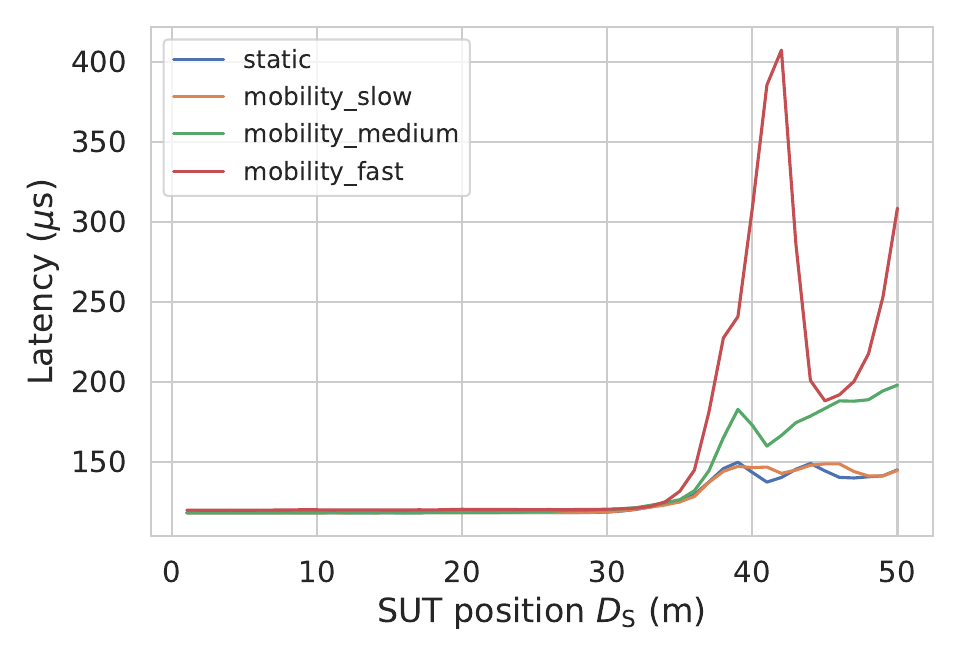}}
    \vspace{-0.4cm}
    \caption{Average latency (NO\_INT configuration).}
    \vspace{-0.2cm}
    \label{fig:NOINT_AvgLat}
\end{figure}
\begin{figure}[t]
    \centerline{\includegraphics[width=0.9\columnwidth]{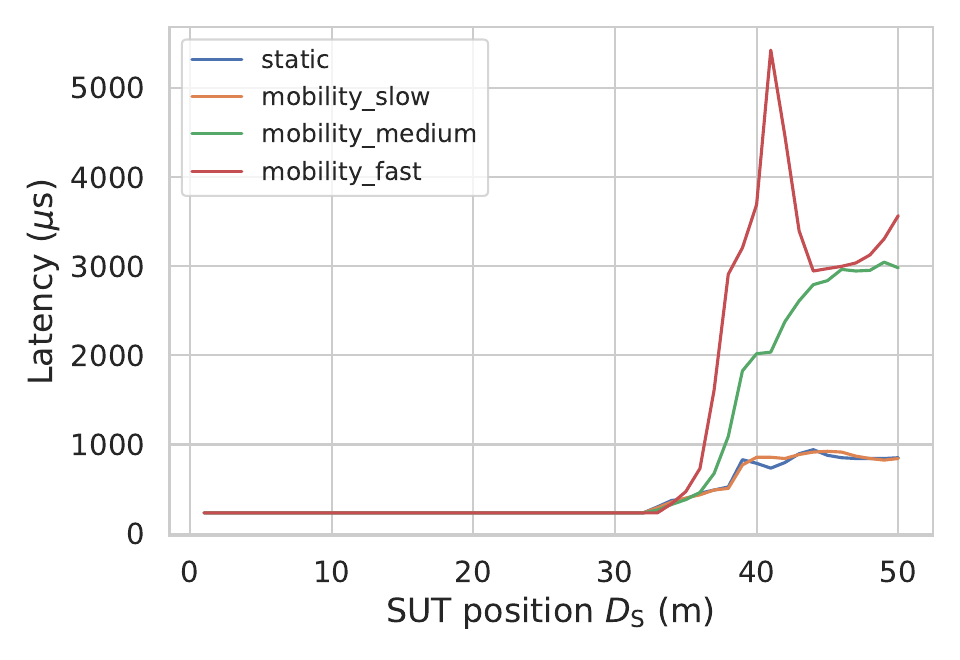}}
    \vspace{-0.4cm}
    \caption{$99$th latency percentile (NO\_INT configuration).}
    \vspace{-0.2cm}
    \label{fig:NOINT_Perc99}
\end{figure}

Results about the NO\_INT configuration, where no interfering nodes are present, are reported in Fig.~\ref{fig:NOINT_AvgLat} and Fig.~\ref{fig:NOINT_Perc99}.
Latency is almost constant at short distances ($D_\mathrm{S}$), while it deteriorates when the SUT is moved farther away.
This is caused by signal attenuation, which leads to re-transmission and to the selection of lower MCS values.
It is interesting to notice that the trend of latency increase is not monotonic with distance.
As described in~\cite{chiavassa2025suitabilitywifiinterconnectingmoving}, this is due to Minstrel behavior: for different SUT positions, the frequency at which each MCS is selected does not always compensate for changes in their failure rate.

What this study wants to point out, however, is the dependence between latency and device speed.
As can be seen in the figures, the behavior of a slowly moving device (\textit{mobility\_slow}) is about the same as the \textit{static} scenario.
Average latency (Fig. \ref{fig:NOINT_AvgLat}) worsens up to around $\SI{200}{\mu s}$ when speed is intermediate (\textit{mobility\_medium}), and can be as high as $\SI{400}{\mu s}$ for a fast-moving device (\textit{mobility\_fast}).
Similar considerations can be made for the $99$th percentile (Fig.~\ref{fig:NOINT_Perc99}), which is noticeably larger than the mean value.

In any case, increasing the movement speed leads to a faster variation of the spectrum conditions observed by the SUT.
Since Minstrel relies on the outcome of past transmission to determine the optimal MCS, which are performed by the application at a fixed pace (once every $\SI{0.5}{s}$), a faster speed implies that decisions of Minstrel are made on outdated information, which explains why, for the same distance $D_\mathrm{S}$, a worse quality of service is obtained.
This is also supported by the fact that significant latency separation between different mobility scenarios only starts at farther SUT positions, where selecting the wrong MCS can lead to retransmissions.
In fact, at shorter distances, any MCS that is selected for transmission has a very high probability of succeeding.

\begin{figure}[t]
    \centerline{\includegraphics[width=0.9\columnwidth]{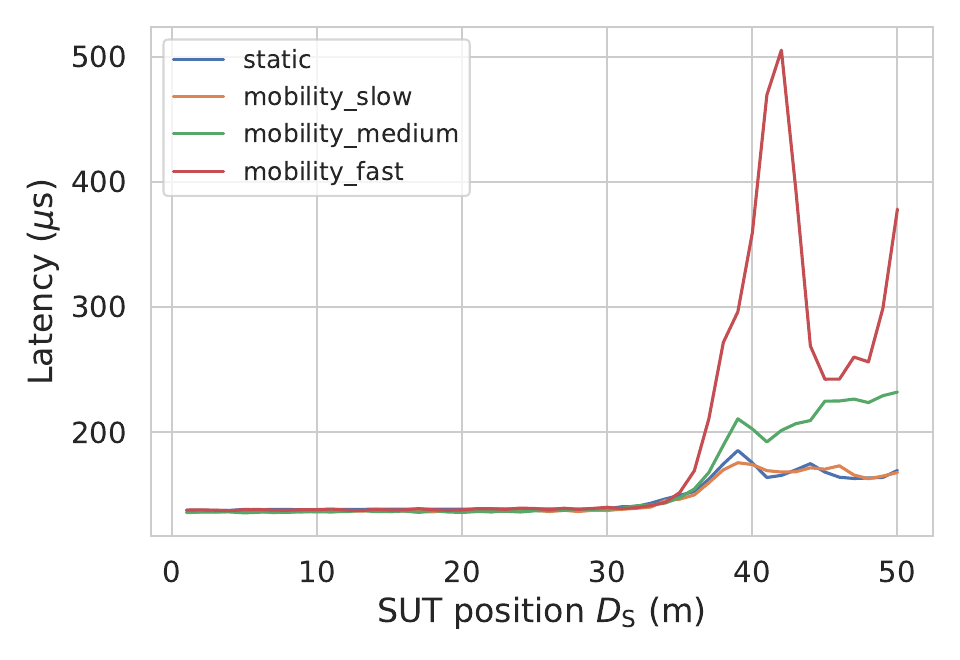}}
    \vspace{-0.4cm}
    \caption{Average latency (VISIBLE configuration).}
    \vspace{-0.2cm}
    \label{fig:VISIB_AvgLat}
\end{figure}
\begin{figure}[t]
    \centerline{\includegraphics[width=0.9\columnwidth]{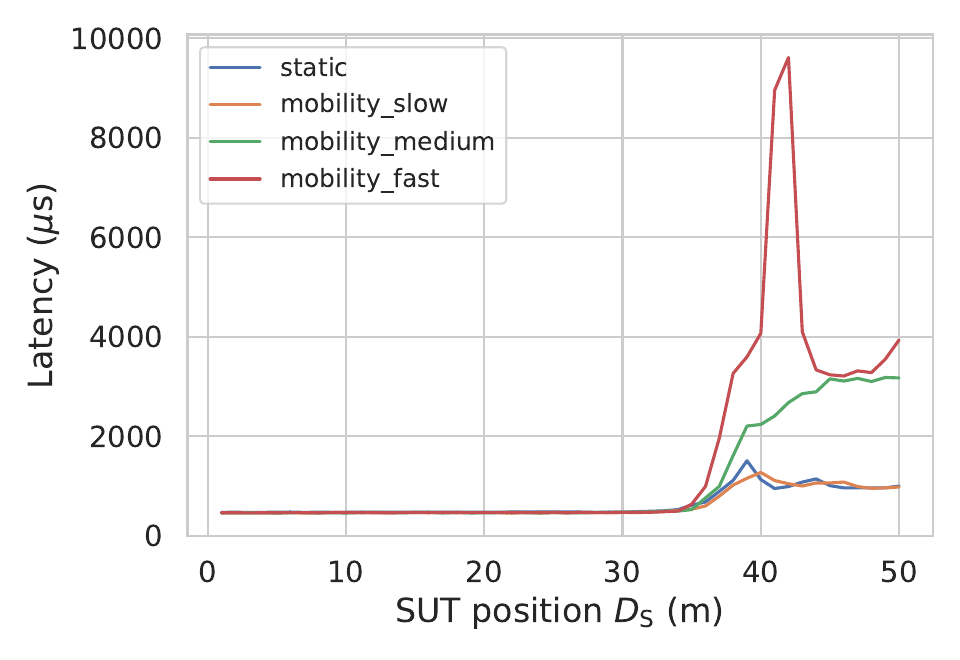}}
    \vspace{-0.4cm}
    \caption{$99$th latency percentile (VISIBLE configuration).}
    \vspace{-0.2cm}
    \label{fig:VISIB_Perc99}
\end{figure}

Figs.~\ref{fig:VISIB_AvgLat} and \ref{fig:VISIB_Perc99}, instead, depict the mean value and the $99$th percentile of latency in the VISIBLE configuration, where the interferer is always in range of both the SUT and the AP.
The behavior mostly resembles the previous case, but the $99$th percentile can now be as high as $\SI{9.6}{ms}$, almost twice as much as NO\_INT.
This is reasonable because the kind of interfering traffic INT injects on air has little impact on average, but can delay tangibly some packets, causing high percentiles to worsen in a noticeable way.
As before, and for the same reasons, increasing the movement speed of the SUT worsens the communication latency.

\begin{figure}[t]
    \centerline{\includegraphics[width=0.9\columnwidth]{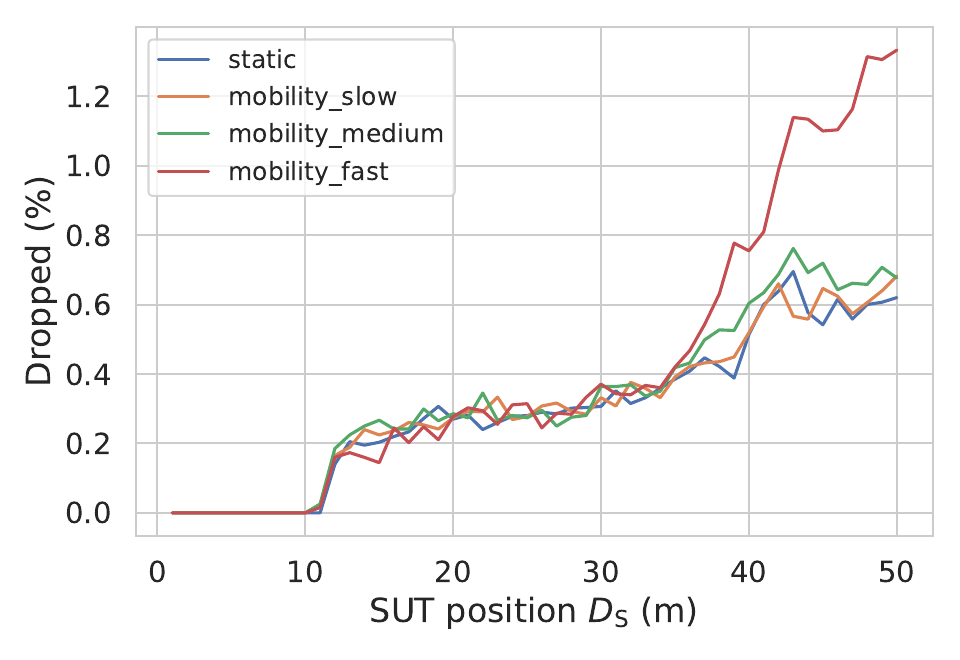}}
    \vspace{-0.4cm}
    \caption{Fraction of dropped packets (HIDDEN configuration).}
    \vspace{-0.2cm}
    \label{fig:HIDDEN_Dropped}
\end{figure}

\begin{figure}[t]
    \centerline{\includegraphics[width=0.9\columnwidth]{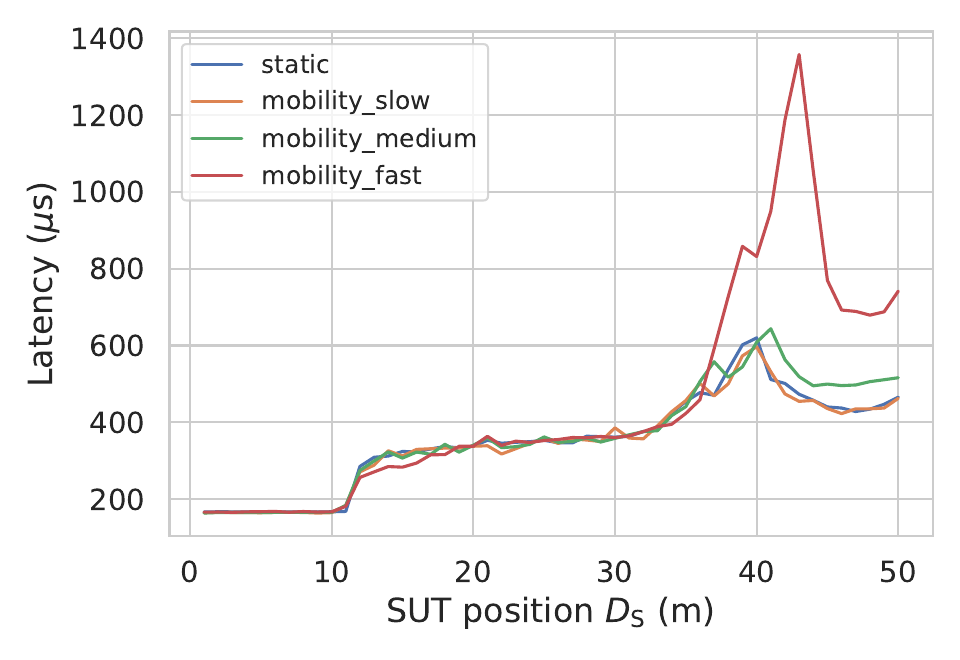}}
    \vspace{-0.4cm}
    \caption{Average latency (HIDDEN configuration).}
    \vspace{-0.2cm}
    \label{fig:HIDDEN_AvgLat}
\end{figure}

\begin{figure}[t]
    \centerline{\includegraphics[width=0.9\columnwidth]{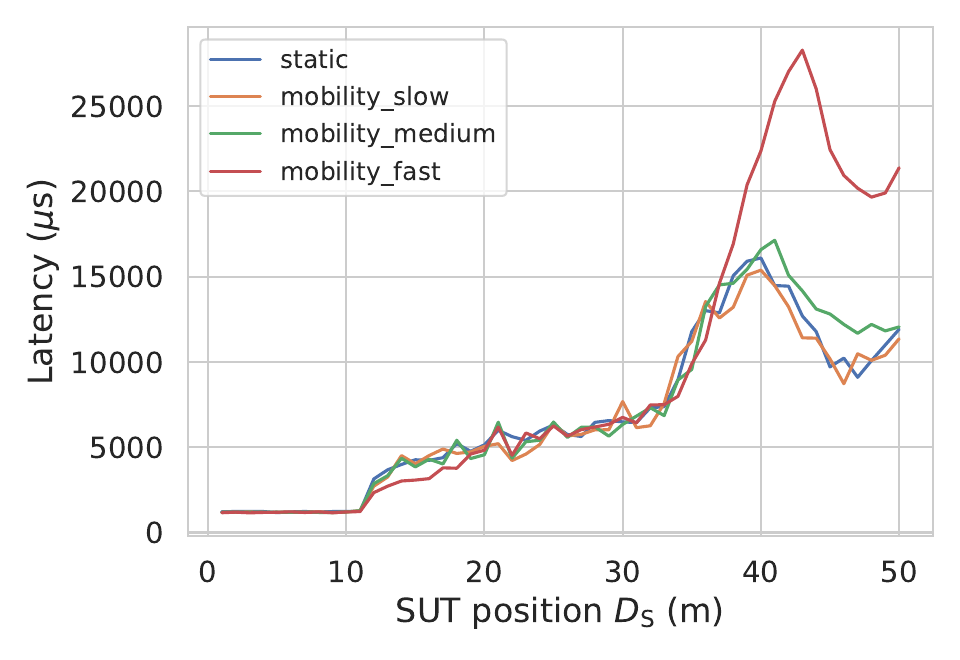}}
    \vspace{-0.4cm}
    \caption{$99$th latency percentile (HIDDEN configuration).}
    \vspace{-0.2cm}
    \label{fig:HIDDEN_Perc99}
\end{figure}

Results for the HIDDEN configuration are reported in Figs.~\ref{fig:HIDDEN_Dropped},~\ref{fig:HIDDEN_AvgLat} and \ref{fig:HIDDEN_Perc99}.
Unlike the NO\_INT and VISIBLE cases, packets are dropped due to reaching the retransmission limit when $D_\mathrm{S}>\SI{11}{m}$.
This is a result of the effect of the hidden node, which impairs the ability of the carrier sense mechanism to prevent concurrent transmissions.
The phenomenon can be observed in Fig. \ref{fig:HIDDEN_Dropped}, where the percentage of dropped packets is plotted as a function of SUT position for every mobility scenario.
Due to the higher number of transmission attempts required to deliver a packet, latency also increases sensibly with respect to the VISIBLE condition.
The $99$th percentile, in this case, can be higher than $\SI{25}{ms}$.

Concerning the dependence between moving speed and latency, as before, separation between the different mobility scenarios can only be observed for farther SUT positions.
The influence of the hidden node, starting at $D_\mathrm{S}>\SI{11}{m}$, does not seem to affect the adaptability of Minstrel when increasing SUT speeds.
Finally, at higher distances, the relative increase of latency w.r.t. the static scenario when mobility is introduced is lower than for the NO\_INT and VISIBLE configuration.
This shows that the effects of the collision avoidance mechanism not operating properly can be more influential than a less accurate MCS selection.
However, as shown in Fig.~\ref{fig:HIDDEN_Dropped}, fast SUT speeds can significantly increase the number of dropped packets.

\section{Conclusion}
\label{sec:Conclusions}
Although not part of the IEEE 802.11 standard, rate adaptation mechanisms like Minstrel are customarily included in almost all Wi-Fi drivers, and specific support for them is provided by the vast majority of Wi-Fi adapters.
By adaptively selecting the best MCS for transmission attempts at any given time, they manage to optimize the quality of service offered to applications operating on a wireless link.

This paper specifically analyzes the effect of node mobility, and, in particular, the speed at which Wi-Fi nodes are moving, considering one of the most important metrics for industrial applications, i.e., transmission latency (both on average and through specific high percentiles).

As results show, increasing node speed impairs the ability of rate adaptation to find the optimal data rate, causing higher latency.
The effect can still be relevant even in the presence of a hidden node.
This opens the possibility of developing methods and strategies, such as network digital twins~\cite{2024-WFCS-SCANZIO}, to inform devices about the best data rate to be used, given their position and the current spectrum conditions.
This is part of our future work on this subject.

\bibliographystyle{IEEEtran}
\bibliography{bibliography}

\end{document}